\documentclass[12pt]{article}
\usepackage{epsf, cite, amssymb}
\usepackage{epsfig}
\usepackage{amsmath}
\makeatletter
\renewcommand\section{\@startsection {section}{1}{\z@}%
                                   {-3.5ex \@plus -1ex \@minus -.2ex}%nn
                                   {2.3ex \@plus.2ex}%
                                   {\normalfont\large\bfseries}}
\renewcommand\subsection{\@startsection{subsection}{2}{\z@}%
                                     {-3.25ex\@plus -1ex \@minus -.2ex}%
                                     {1.5ex \@plus .2ex}%
                                     {\normalfont\bfseries}}
\makeatother

\let\non\nonumber

\newcommand{\bea}{\begin{eqnarray}}
\newcommand{\eea}{\end{eqnarray}}
\newcommand{\be}{\begin{equation}}
\newcommand{\ee}{\end{equation}}

\newcommand{\hlf}{\frac{1}{2}}

\newcommand{\Z}{{\mathbb Z}}
\newcommand{\R}{{\mathbb R}}

\newcommand{\T}{\theta}
\newcommand{\G}{\Gamma}

\newcommand{\e}{\epsilon}

\newcommand{\dd}{\delta}

\newcommand{\rr}{\rightarrow}
\newcommand{\m}{\mu}
\newcommand{\n}{\nu}
\newcommand{\p}{\partial}

\newcommand{\tz}{\tilde z}
\newcommand{\tg}{\tilde g}

\newcommand{\hz}{\hat z}
\newcommand{\hx}{\hat x}
\newcommand{\hxm}{\hat{x}^-}
\newcommand{\hxp}{\hat{x}^+}

\newcommand{\tK}{\widetilde{K}}

\newcommand{\xp}{x^+}
\newcommand{\xm}{x^-}
\newcommand{\s}{\sigma}
\renewcommand{\H}{\mathcal{H}}
\newcommand{\La}{\Lambda}
\newcommand{\la}{\lambda}
\newcommand{\Tr}{\operatorname{Tr}}
\newcommand{\U}{\operatorname{U}}
\newcommand{\MC}{\mathbb{C}}
\newcommand{\ap}{\alpha'}
\newcommand{\w}{\wedge}

\newcommand{\C}[1]{$(\ref{#1})$}

\begin{document}
\begin{titlepage}

\begin{center}

{September 26, 2005}
\hfill         \phantom{xxx}

\hfill EFI-05-11

\vskip 2 cm
{\Large \bf A Matrix Model for the Null-Brane}\\
\vskip 1.25 cm { Daniel Robbins\footnote{email address:
robbins@theory.uchicago.edu} and Savdeep Sethi\footnote{email address:
 sethi@theory.uchicago.edu}}\\
{\vskip 0.5cm  Enrico Fermi Institute, University of Chicago,
Chicago, IL
60637, USA\\}
%\vskip 0.2cm $^b$ School of Natural Sciences, Institute for
%Advanced Study, Princeton, NJ 08540}

\end{center}

\vskip 2 cm

\begin{abstract}
\baselineskip=18pt
The null-brane background is a simple smooth $1/2$ BPS solution of 
string theory. By tuning a parameter, this background develops a big 
crunch/big bang type singularity. We construct the DLCQ 
description of this space-time in terms of a Yang-Mills theory on a time-dependent space-time. Our dual Matrix description provides a non-perturbative framework in which the fate 
of both (null) time, and the string S-matrix can be studied. 

\end{abstract}

\end{titlepage}

\pagestyle{plain}
%\baselineskip=18pt
% Try a wider skip
\baselineskip=19pt

\section{Introduction}

Understanding the physics of the big bang is one of the key questions
facing string theory. Past work on cosmological singularities suggests
that perturbative string theory breaks down near the 
singularity~\cite{Liu:2002ft, Liu:2002kb, Lawrence:2002aj, Horowitz:2002mw, Berkooz:2002je, Cornalba:2003kd}. See~\cite{related, Simon:2002ma, Elitzur:2002rt, Fabinger:2002kr, Berkooz:2005ym} for some related work. What is needed is a different formulation of 
physics in the regime of strong gravity near the singularity, 
perhaps via holography. 

Such dual descriptions, in the spirit of AdS/CFT, have been studied 
in~\cite{Elitzur:2002rt, holo, holo2}. Recently, dual descriptions of the light-like
linear dilaton and related solutions have been described 
in~\cite{Craps:2005wd,Li:2005sz, Li:2005ti, Kawai:2005jx, Hikida:2005ec, Das:2005vd, Chen:2005mg, She:2005mt, Chen:2005bk, Ishino:2005ru}\ via Matrix 
theory~\cite{Banks:1996vh}. These backgrounds always contain a 
region with a cosmological singularity where perturbative string theory
breaks down.    

The aim of this work is to extend these ideas to the null-brane solution. 
The null-brane is constructed as a
quotient of flat space, $\R^{1,3}$. The quotient action is generated by an element
of the Poincar\'e group containing a boost, a rotation and a
shift. When viewed as a quotient space, the metric is flat. However, when 
expressed in more natural coordinates, the resulting metric is not flat 
but generalizes the flux-brane solutions
corresponding to Melvin universes. 
Instead of just a magnetic field (as in
the Melvin case), there are both electric and magnetic fields. This
class of space-time is therefore a sort of 
Melvin universe with electric
fields. In~\cite{Figueroa-O'Farrill:2001nx,Simon:2002ma}, this space-time 
was termed a ``null-brane.'' 

\begin{figure}[h] \label{figure1}
\begin{center}
\includegraphics[width=10cm]{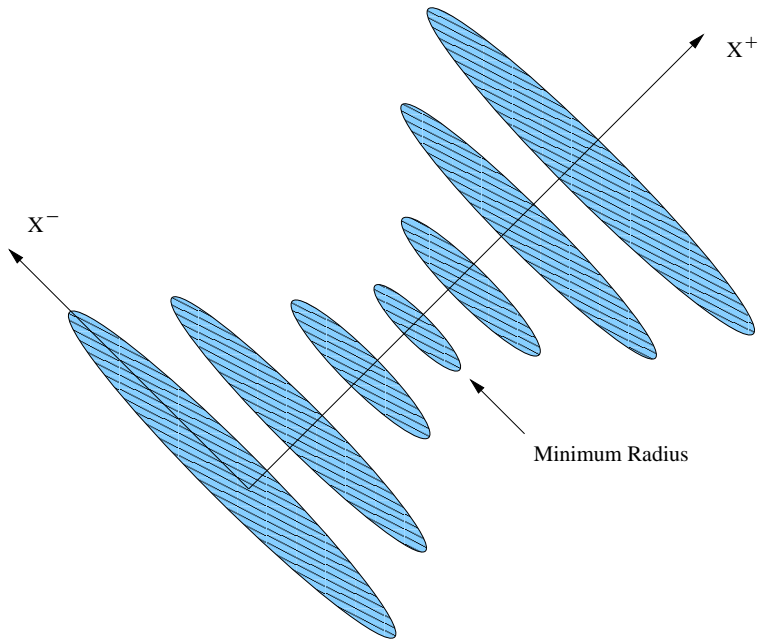}
\caption{The circle radius shrinks to a minimum $L$ at $\xp =0$.}
\end{center}
\end{figure}

The basic structure of the space-time is depicted in figure
\ref{figure1}. There is a circle whose radius shrinks as we increase
$\xp$ until it reaches size $L$ at $\xp=0$. The size, $L$, is a tunable
modulus in the metric. Viewing $\xp$ as
light-cone time, we see that a particle becomes blue-shifted as
time evolves by an amount that increases with decreasing $L$. The
singular limit corresponds to taking $L\rightarrow 0$. The resulting
singular space has been considered in~\cite{Liu:2002ft}. From this
perspective, the background is light-cone time-dependent. 

The aim of this work is to find a DLCQ description of the 
null-brane. We should note that for non-zero $L$ this
space-time has the following virtues. First, there are no pathologies: 
neither curvature singularities nor closed causal curves. Second, there 
is a null killing vector which facilitates string 
quantization. A space-time with these properties serves as a good 
perturbative string background with an S-matrix. Indeed, string scattering
has been studied on this 
background~\cite{us, Liu:2002kb, Fabinger:2002kr}.  
However, on taking the limit $L \rr 0$, the space-time develops a 
null singularity. This is an added
feature that allows us to access the physics of a big crunch/big bang
singularity in what we might hope is a controlled manner. 

In section $2$, we define the null-brane quotient and study M-theory and string theory compactified on this background at the level of supergravity. In section $3$, we describe a decoupling limit that captures the 
DLCQ physics of the null-brane. In section $4$, we derive the Matrix description of M-theory on the null-brane for the case $N=1$ of a single D-brane using the DBI action. This model is already quite fascinating: it looks like a $1+1$-dimensional field theory on a cylinder whose radius is time-dependent. In the far past and the far future, the cylinder shrinks to zero size. The cylinder reaches a maximum radius at $x^+=0$ proportional to $1/L$ which diverges as $L \rr 0$. This is quite reminiscent of the way in which Milne space appeared as the string worldsheet in~\cite{Craps:2005wd}. It should be contrasted with the holographic description of branes wrapping the null-brane which gives a space-time-dependent non-commutative field theory~\cite{holo}.  

We then proceed to conjecture the complete non-abelian answer for many branes using
results based in part on the quotient description of the null-brane and in part on the DBI approach. We then present some additional arguments suggesting that our final Matrix Lagrangian is complete.

\section{Defining the Background}
\subsection{The orbifold group}
\label{orbifoldgroup}

We define our background as follows: consider $\R^{1,3}$
parametrized by coordinates $$x^\pm = \frac{1}{\sqrt 2}(x^0 \pm
x^1), \, x, \, z, $$ with the usual metric $ds^2 = -2dx^+ dx^- +
dx^2 + dz^2$. We act on these coordinates by an element of the
4-dimensional Poincar\'e group: \be \label{gen} g = \exp(2\pi i
K);\quad K = \frac{\lambda}{\sqrt 2}(J^{0x} + J^{1x}) + L P^z, \ee
where $L$ has dimensions of length. This is the only scale
beyond the Planck scale in our setup. Under this action which depends on $(\lambda, L)$, \be
\label{quotgroup}
 X =
\left(\begin{array}{c}x^+\\x^-\\x\\z\\\end{array}\right)\
\rightarrow \quad
g\cdot X = \left(\begin{array}{c}x^+\\x^- + 2\pi \lambda x + 2\pi^2 \lambda^{2} x^+\\x + 2\pi \lambda x^+\\
z + 2\pi L\\\end{array}\right). \ee 
The parameter $\lambda$ can be set to one by a light-cone boost
\be\label{redef}
x^{+} \rr {x^{+} \over \lambda}, \quad x^{-} \rr \lambda x^{-}. 
\ee
{}For most of our discussion, we will assume $\lambda=1$ except when we discuss decoupling in 
section~\ref{decoupling}.
The length squared of
closed curves can be easily computed, \be \left(g^n\cdot X -
X\right)^2 = (2\pi nx^+)^2 + (2\pi nL)^2 > 0\ . \ee There are
no closed causal curves. For sufficiently low energy scattering, 
we therefore need not worry about
effects from large back reaction invalidating perturbative string
computations.

It is worth stressing that four of the ten Poincar\'e generators are unbroken -- those that
commute with $K$.  These are 
$$P^+, \quad P^z, \quad K,\quad \tK =
\frac{1}{\sqrt 2}(J^{0z} + J^{1z}) + L P^x. $$ 
A quotient group
element $g$ acts on the momenta in the following way:
 \be P=\left(\begin{array}{c}p^+\\p^-\\p^x\\p^z\\\end{array}\right);\quad
g\cdot P = \left(\begin{array}{c}p^+\\p^- + 2\pi p^x + 2\pi^2 p^+\\
p^x + 2\pi p^+\\p^z\\\end{array}\right). \ee We note that for
this orbifold, it is not the case that any point with $x^+<0$ is
in the causal past of every point with $x^+>0$. To check this, we
compute \be  \begin{array}{ccl}
 \left(X - g^n\cdot\tilde X\right)^2 &=& -2\Delta x^+ \Delta x^- +
(\Delta x)^2
+ (2\pi n)^2 x^+ \tilde x^+ + 2(2\pi n)(x^+ \tilde x - x\tilde x^+) \\
&& + (\Delta z)^2 - 2(2\pi n)L\Delta z + (2\pi n)^2
L^2, \end{array} \ee where $\Delta x^\mu = x^\mu - \tilde
x^\mu$. At large $n$, we have $(2\pi n)^2(x^+\tilde x^+ +
L^2)$, so only points with $x^+ \tilde x^+ < -L^2$ are
always causally related in this way.

The orbifold action lifts to the spin bundle over $\R^{1,3}$. To
determine the number of preserved supersymmetries, we need to
count the number of spinors, $\e$, left invariant by (the lift
of) $K$. The $P^z$ term in $K$ does not act on a spinor. 
In terms of standard real gamma matrices, $\G^\m$, satisfying the
Clifford algebra relations, 
$$ \{ \G^\m, \G^\nu \} = \eta^{\mu\nu}, \quad \m,\nu =0,\ldots, 10,$$
it is
easy to check that the invariance condition, \be \label{qsusy}
\left(\Gamma^{0x} + \Gamma^{1x} \right) \e =0, \ee implies that
\be \Gamma^+ \e =0. \ee
This background therefore
preserves one-half of the available supersymmetry. To construct a
string or M-theory background, we simply append an additional flat $\R^6$ 
or $\R^7$ factor to give a $10$ or $11$-dimensional metric.

\subsection{The null-brane background}
\label{nullmetric} 

It is natural to express the metric in terms of new variables in which the
quotient action simplifies. This choice of coordinates makes it easy
to reduce along orbits of $K$. Let us
perform the following change of variables: \be \label{hatted}
\hat x^+ = x^+ ,\quad \hat x^- = x^- - {z x \over L} + { z^2
x^+ \over 2 L^2} ,\quad \hat x = x - {z x^+ \over L},
\quad \hat z = \frac{z}{L}. \ee The hatted $x$-coordinates
are natural because they are invariant under the action of $K$.
The group element $g$ of equation \C{gen}\ acts only by
translation on $\hz$ sending $$\hat z \rightarrow \hz +2\pi.$$
In these coordinates, the metric takes the form
\be\label{invmetric}
ds^2 = -2 d \hat x^+ d \hat x^- + d \hat x^2 +
( (\hxp)^2 + L^2) d \hat z^2 + {2} (\hat x^+ d \hx - \hat x d \hat x^+)
 d \hat z.
\ee
This metric was obtained by a coordinate change from flat space so there
is no curvature. 

\subsection{M-theory on the null-brane}

Let us consider M-theory on this space-time and reduce to type IIA on
the $ \hat z$ circle. 
We obtain a solution similar to a
flux-brane, but with a null RR 1-form field strength. 
After massaging \C{invmetric}\ into the standard form for determining
the string metric, we see that the flat
11-dimensional metric becomes: 
\be \label{11metric}
\begin{array}{ccl} ds_{11}^2 &=&
-2dx^+ dx^- + dx^2 + dz^2 + (ds_7)^2 \\
&=& -2d\hxp d\hxm - {\hx^2\over \Lambda} (d\hxp)^2 +  {2 \hx
\hxp\over \Lambda} d\hx d\hxp + \Lambda \left( d \hz + {\hxp \over
\Lambda} d\hx - {\hx \over \Lambda} d\hxp \right)^2 \\ & & +
{L^2\over \Lambda}
d\hx^2 + (ds_7)^2 \\
\end{array}
\ee where \be \Lambda = (\hat x^+)^2 + L^2. \ee To obtain
the string frame metric, we use the usual relation \be ds_{11}^2
= e^{4\phi/3} (d\hz + A)^2 + e^{-2\phi/3} ds_{10}^2 \ee where
$ds_{10}^2$ is the string frame metric, and $A$ is the RR 1-form.
Using this relation, we read off the following string metric,
dilaton, and RR 1-form potential: \bea ds_{10}^2 &=&
\Lambda^{1/2} \left\{ -2d\hxp d\hxm - {\hx^2\over \Lambda}
(d\hxp)^2 +  {2 \hx \hxp\over \Lambda} d\hx d\hxp +
{L^2\over \Lambda} d\hx^2 + (ds_7)^2 \right\}
\\ \phi &=& {3\over 4} \, {\rm log}\, \Lambda \label{stringmetric}\\
A &=& \left( {\hxp} d\hx - {\hx} d\hxp \right)/\Lambda. \eea The
field strength $F$ associated to $A$ is null, $$ F =
{2L^2\over \Lambda^2} \, d\hxp \wedge d\hx, $$ which is the
reason for the terminology ``null-brane'' given
in~\cite{Figueroa-O'Farrill:2001nx}. Note that the string coupling becomes small at $\hxp=0$ when 
we take $L \rr 0$.  

\subsection{Type II string theory on the null-brane}

We now turn to type II string theory quotiented by the
action~\C{quotgroup}, or equivalently with the metric~\C{invmetric}.
We have no $B$-field and no RR fields.  The dilaton is
constant, $g_s = e^{\Phi_0}$. What happens as
$L$ becomes small compared to the string scale? It seems wise to see what duality at the level of
the supergravity solution can teach us. 

In the limit where
$L\rightarrow 0$, the metric develops a singularity at $\hxp=0$ which
is basically the $\hz$ circle shrinking to zero size resulting in a closed
null curve. It is
natural to therefore T-dualize along $\hz$ which results in the metric 
(see, for example,~\cite{Bergshoeff:1995as})
\bea \label{tdualmet}
ds_{\mathrm{T-dual}}^2 &=& -2dx^+ dx^- - \frac{x^2}{L^2 + (x^+)^2} (dx^+)^2 + 
2\frac{x^+ x}{L^2 +  (x^+)^2} dx^+ dx \cr & & + \frac{L^2}{L^2 + (x^+)^2} dx^2
+ \frac{1}{L^2 + (x^+)^2} d\tz^2
\eea
where the T-dual coordinate $\tz$ still has a period of $2\pi$ (we use units where 
$\alpha' = 1$ for the moment). We have dropped the hats for the T-dual
variables. There are also $B$-fields generated
\bea \label{fluxes}
B_{+\tz} &=& \frac{x}{L^2 + (x^+)^2}, \\
B_{\tz x} &=& \frac{x^+}{L^2 + (x^+)^2}, \non
\eea
and the dilaton is no longer constant,
\bea\label{dilaton}
\Phi &=& \Phi_0 - \hlf\ln\left(L^2 + (x^+)^2 \right), \non\\
 \tg_s &=& \frac{g_s}{\sqrt{L^2 + (x^+)^2}}. 
\eea
The first thing to note is that if we hold $g_s$ fixed and take
$L\rightarrow 0$, the dual coupling diverges at $x^+=0$. From the
original quotient group perspective, this corresponds to going over to the
parabolic orbifold studied in \cite{Horowitz:1990bv, Liu:2002ft}.   

The $B$-field gives a field strength whose only non-vanishing component is
\be\label{hfield}
H_{+x\tz} = -\frac{2L^2}{\left(L^2 + (x^+)^2 \right)^2}. 
\ee
This field strength diverges as $L\rr 0$ at $x^+ \rr 0$. This is
intriguing and suggests the existence of a kind of critical theory of
closed strings in a large light-light $B$-field. There is a strong analogy with open
strings in a light-like constant 2-form field strength, and there might well be a relation with the non-relativistic strings studied in~\cite{Gomis:2005pg}. 
The metric~\C{tdualmet}\ is now curved with non-vanishing curvature components:
\bea
& R_{+\tz +}\vphantom{R}^{\tz} = \frac{L^2 - 2(x^+)^2}{
\left(L^2 + (x^+)^2 \right)^2},  \qquad \qquad
& R_{+x+}\vphantom{R}^x = \frac{3L^2}{\left(L^2 + (x^+)^2 \right)^2}, \non\\
& R_{+x+}\vphantom{R}^- = \frac{3L^2 x^+ x}{\left(L^2 + (x^+)^2
  \right)^3},
\qquad\qquad 
& R_{+xx}\vphantom{R}^- = \frac{3L^4}{\left(L^2 + (x^+)^2 \right)^3}, \non\\
& R_{++} = \frac{4L^2 - 2(x^+)^2}{\left(L^2 + (x^+)^2 \right)^2}, \qquad\qquad
& R = 0. 
\eea
It is not hard to check that this dilaton, $H$-field and Ricci tensor
combine to give a good string background with vanishing beta
functions as we expect. It is also worth noting that as $L \rr 0$ with $x^+ \gg L$, 
$H \rr 0$, but the string coupling and curvature are still nontrivial:
\be
\tg_s \rr {g_s \over |x^+|}, \qquad\qquad R_{++} \rr -{2\over (x^+)^2}.
\ee

Finally, we would like to lift this configuration to M-theory. This is natural if 
we consider type IIA on the metric~\C{tdualmet}, and we choose to hold $g_s$ fixed 
but consider $L\rr 0$. Let $y$ denote the coordinate of the M-theory circle, we
then obtain the
following $11$-dimensional solution:
\bea
ds_{11}^2 &=& (g_s)^{-\frac{2}{3}} \left\{L^2 + (x^+)^2 \right\}^{\frac{1}{3}}
ds_{\mathrm{T-dual}}^2 + (g_s)^{\frac{4}{3}} 
\left\{L^2 + (x^+)^2 \right\}^{-\frac{2}{3}} dy^2, \\
&=& \left(g_s \{L^2 + (x^+)^2 \} \right)^{-\frac{2}{3}} \left[
-2\left\{L^2 + (x^+)^2 \right\} dx^+ dx^- - x^2 (dx^+)^2 \right.\non\\
&& \left.\vphantom{\left(L^2 + (x^+)^2 \right)} + 2x^+ x dx^+ dx + L^2 dx^2 
+ d\tz^2 + g_s^2  dy^2 \right], \non
\eea
and a longitudinal 3-form potential,
\bea
C_{+\tz y} &=& \frac{2}{3} \frac{x}{L^2 + (x^+)^2}, \\
C_{\tz xy} &=& \frac{2}{3} \frac{x^+}{L^2 + (x^+)^2}, \non
\eea
with 4-form field strength
\be
G_{+x\tz y} = -\frac{4}{3}\frac{L^2}{\left\{L^2 + (x^+)^2 \right\}^2}.
\ee
The curvature of the metric can be computed.  We will quote only the Ricci
tensor whose non-vanishing component is
\be
R_{++} = \frac{8L^2 - 12(x^+)^2}{\left\{ L^2 + (x^+)^2 \right\}^2}.
\ee
The Ricci scalar vanishes as before.  Lastly, note that had we considered type IIB on~\C{tdualmet}, 
it would have been natural to use S-duality when the coupling becomes large.

\section{The DLCQ Description} 

\subsection{Light-like to space-like compactification}

We first note that the action of $p^+$ commutes with the null-brane quotient~\C{quotgroup}. This means we 
can compactify the  $\hxm$ direction, 
\be
\hxm \sim \hxm + R, 
\ee
and consider the sector with fixed light-cone momentum ${\hat p}^+ = N/R$. 

We cannot relate this light-like compactification to a space-like compactification using the procedure of~\cite{Seiberg:1997ad}\
because the metric~\C{invmetric}\ depends explicitly on $\hxp$. However, we can use the modified procedure of~\cite{Craps:2005wd}. Choose a direction ${\hat x}^1$ and make the identifications
\be
\left(\hxp, \hxm, {\hat x}^1 \right) \sim \left(\hxp, \hxm, {\hat x}^1 \right) + \left(0, R, \e R \right).
\ee
The Lorentz transformation
\be \label{Lorentz}
\hxp = X^+, \qquad
\hxm = {X^+ \over 2\e^2} + {X^-} + { {X}^1 \over \e}, \qquad
{\hat x}^1 = {X^+ \over \e} + X^1,  \qquad \hx^i  = X^i \quad i>1, 
\ee
while holding fixed $L$ and
\be
z = {Z},  
\ee
results in the M-theory metric
\be\label{mmetric}
ds_{11}^2  = -2 dX^+ dX^- + dX^2 + \left( (X^+)^2 + L^2 \right) dZ^2 + 2 \left( X^+ dX - X dX^+ \right) dZ + \sum_{i=1}^7 (dX^i)^2 
\ee
with the identifications 
\be\label{slightly}
Z \sim Z + 2\pi, \qquad X^1 \sim X^1 + \e R. 
\ee
There are $N$ units of momentum in the $X^1$ direction. 

We reduce to type IIA on the $X^1$ circle. This is a straightforward reduction which leaves us with type IIA
on a space with metric
\be
ds_{10}^2  = -2 dX^+ dX^- + dX^2 + \left( (X^+)^2 + L^2 \right) dZ^2 + 2 \left( X^+ dX - X dX^+ \right) dZ + \sum_{i=2}^7 (dX^i)^2 
\ee
and $N$ D0-branes. This is the theory of $N$ D0-branes on the 
null-brane quotient.

We can also arrive at this same conclusion by directly studying the orbifold action~\C{quotgroup}. It is easy to check that the identification
\be \label{twisted}
\left(\xp, \xm, {x}^1 \right) \sim \left(\xp, \xm, {x}^1 \right) + \left(0, R, \e R \right)
\ee
commutes with the orbifold action. After making the same Lorentz transformation given in~\C{Lorentz}, the DLCQ identification becomes
\be\label{spacelike}
X^1 \sim X^1 + \e R. 
\ee 
Using either approach, we reduce the study of the light-like compactified null-brane in M-theory to the study of the dynamics of $N$ D0-branes on the null-brane quotient.  Our task in section~\ref{dbranes}\ is to determine this theory. 

\subsection{A decoupling limit}
\label{decoupling}

Note, however, that this procedure does {\it not} result in a decoupling limit because the transformation~\C{Lorentz}\ does not involve a rescaling of $\hat{x}^+$ so the corresponding light-cone energy $\hat{p}^-$ does not become small.

To obtain a decoupling limit, we need to perform an additional transformation. Let us return to the orbifold description~\C{quotgroup}\ with $\lambda$ a free parameter.  First note that the identification~\C{twisted}\ implies that
\be \label{relationship}
p^{+} = \e p^{1}
\ee 
if we stay in the DLCQ sector with fixed $N$. 

The Lorentz transformation~\C{Lorentz}\ applied to the flat space variables takes us to a space-like circle but does not scale the light-cone energy $E^{-}$. Rather the energy and momenta transform in the following way 
\be
E^{-} \rr E^{-} + {p^{+} \over 2 \e^{2}} - {p^{1}\over \e}, \qquad p^{+}\rr p^{+}, \qquad p^{1} \rr p^{1} - {p^{+}\over \e}.
\ee
The mass shell condition,
\be
-2 E^{-} p^{+} + p^{i} p_{i} + m^{2}=0,
\ee
together with the relation~\C{relationship}\ implies that $p^{i} \sim O(\e)$ while $p^{+} \sim O(\e^{2})$ and $E^{-} \sim O(1)$. The null-brane quotient determined 
by $\lambda$ is unchanged but $X^{1}$ satisfies the condition~\C{spacelike}. 

We now boost to rescale our energies sending
\be
X^{+} \rr \e X^{+}, \quad X^{-}\rr {X^{-}\over \e}. 
\ee
This has the effect of sending $\lambda \rr \e \lambda$ while leaving $L$ invariant. All energies and momenta are now of order 
$\e$. Reducing to type IIA string theory on $X^{1}$ gives us type IIA string theory with 
\be \label{scales}
g_{s} \sim \e^{3/2}, \quad \ell_{s} \sim \e^{-1/2}
\ee
and a flat metric quotiented by the null-brane identification with parameters $ ( \lambda, L)$ where $\lambda \sim \e$. 

We can now change to invariant coordinates using~\C{hatted}\ now including factors of $\lambda$. It is easy to find the resulting 
metric 
\be\label{newmetric}
ds^2 = -2 d \hat x^+ d \hat x^- + d \hat x^2 +
( \{ \lambda \hxp \}^2 + L^2) d \hat z^2 + {2 \lambda} (\hat x^+ d \hx - \hat x d \hat x^+)
 d \hat z + d\hx^i d\hx^i.
\ee
By rescaling $\hz$, we see that this metric really depends on the combination $L/\e\lambda$. For the moment, however, we
choose to keep $\hz$ dimensionless with canonical period $2\pi$.  These scalings define a decoupling limit for M-theory on the null-brane quotient. String oscillators decouple because our characteristic energy $\e E^{-}$ is much smaller than the string scale given in~\C{scales}. Closed strings also decouple because the 10-dimensional Newton constant is becoming small at these energies, 
$$ g_{s}^{2} \left( {\e E^{-} \ell_{s}}\right)^{8} \rr 0, $$
as $\e \rr 0$.  We will apply these scalings to the theory of D0-branes on the null-brane in the following section.

\section{D-branes on the null-brane}
\label{dbranes}

\subsection{Decoupling the DBI action}

An analysis of boundary states in the null-brane appears in~\cite{Okuyama:2002pc}.
Our goal in this section is to derive the gauge theory describing the dynamics of 
$N$ D0-branes on the null-brane. The natural approach to use is the orbifold
 description of the null-brane given by the identification~\C{quotgroup}. This 
 turns out to be subtle for reasons we will describe later. Therefore, we first consider the abelian case with $N=1$ where we can use the DBI action. 
 
We start with type IIA string theory with a single D0-brane moving on a space-time with metric~\C{newmetric}\ where, for the moment, we do not decouple.
A T-duality along $\hz$ converts the D0-brane to a D-string wrapped along the T-dual direction $z$. On performing this T-duality, we find
\bea\label{DBIback}
ds^2 &=& -\frac{\la^2x^2}{\La}(dx^+)^2-2dx^+dx^-+\frac{2\la^2x^+x}{\La}dx^+dx+\frac{L^2}{\La}dx^2+\frac{(\ap)^2}{\La}dz^2+dx^idx^i,\non\\
B &=& \frac{2\ap}{\La}\left(-xdx^++x^+dx\right)\w dz,\non\\
e^{2\Phi} &=& \frac{\ap g_s^2}{\La},
\eea
where
\be
\La=L^2+\la^2(x^+)^2,
\ee
and $z$ is again dimensionless with period $2\pi$.

The DBI action is given by
\be
S=-\frac{1}{\ap}\int d\tau d\s e^{-\Phi}\sqrt{-\det\left[\left(G_{\m\n}+B_{\m\n}\right)\p_a X^\m\p_bX^\n+2\pi\ap F_{ab}\right]}.
\ee
Evaluating this action on the solution~\C{DBIback}\ gives
\bea\label{largeexp}
&S =& -\frac{1}{g_s(\ap)^\frac{3}{2}}\int d\tau d\s \left\{-\La\left[\left(-\frac{\la^2x^2}{\La}(\dot x^+)^2-2\dot x^+\dot x^-+\frac{2\la^2x^+x}{\La}\dot x^+\dot x+\frac{L^2}{\La}\dot x^2+\frac{(\ap)^2}{\La}\dot z^2+\dot x^i\dot x^i\right)\right.\right.\non\\
&& \left.\left.\times\left(-\frac{\la^2x^2}{\La}(x^{+\prime})^2-2x^{+\prime}x^{-\prime}+\frac{2\la^2x^+x}{\La}x^{+\prime}x'+\frac{L^2}{\La}(x')^2+\frac{(\ap)^2}{\La}(z')^2+x^{i\prime}x^{i\prime}\right)\right.\right.\non\\
&& \left.\left.-\left(-\frac{\la^2x^2}{\La}\dot x^+x^{+\prime}-\dot x^+x^{-\prime}-\dot x^-x^{+\prime}+\frac{\la^2x^+x}{\La}\dot x^+x'+\frac{\la^2x^+x}{\La}\dot xx^{+\prime}+\frac{L^2}{\La}\dot xx'+\frac{(\ap)^2}{\La}\dot zz'+\dot x^ix^{i\prime}\right)^2\right.\right.\non\\
&& \left.\left.+\left(-\frac{\la\ap x}{\La}\dot x^+z'+\frac{\la\ap x}{\La}\dot zx^{+\prime}+\frac{\la\ap x^+}{\La}\dot xz'-\frac{\la\ap x^+}{\La}\dot zx'+2\pi\ap F\right)^2\right]\right\}^\hlf.
\eea
Note that a prime denotes a $\sigma$ derivative while a dot denotes a $\tau$ derivative.
We now make the gauge choice 
\be\label{defb}
z=\s
\ee 
so $\s$ has period $2\pi$. The action~\C{largeexp}\ simplifies to:
\bea
&S =-\frac{1}{g_s(\ap)^\frac{3}{2}}\int d\tau d\s &\big\{(\ap)^2\big[\left(2\dot x^+\dot x^--\dot x^2-\dot x^i\dot x^i\right)+4\pi\la F\left(x\dot x^+-x^+\dot x\right) \cr
&&-4\pi^2\La F^2\big]+\ldots\big\}^\hlf, 
\eea
where the dots represent terms that are either sixth order in the $x^\m$, or are $L^2$ times something fourth order in $x^\m$, each with precisely two $\tau$ derivatives and two $\s$ derivatives.

Next we use our gauge freedom to set 
\be \label{defc} x^+=c\tau/\sqrt 2 \ee
 where $c$ is a constant. We expand around the static configuration $x^+ = x^- = c\tau/\sqrt 2$ by substituting $x^-=c\tau/\sqrt 2+\sqrt 2y$ where $y$ is a fluctuation.  The result is
\bea
S &=& -\frac{1}{g_s(\ap)^\frac{3}{2}}\int d\tau d\s\left\{(\ap)^2\left[\left(c^2+2c\dot y-\dot x^2-\dot x^i\dot x^i\right)+2\sqrt 2\pi\la cF\left(x-\tau\dot x\right)-4\pi^2\La F^2\right]\right.\non\\
&& \left.+\hlf\la^2c^2x^2(x')^2+L^2c^2(x')^2+2L^2c\dot y(x')^2-L^2\dot x^i\dot x^i(x')^2+\hlf\la^2c^2x^2x^{i\prime}x^{i\prime}\right.\\
&& \left.+c^2\La x^{i\prime}x^{i\prime}+2c\La\dot yx^{i\prime}x^{i\prime}-\la^2c^2\tau x\dot xx^{i\prime}x^{i\prime}-L^2\dot x^2x^{i\prime}x^{i\prime}-\La\dot x^i\dot x^ix^{j\prime}x^{j\prime}+c^2\La(y')^2\right.\non\\
&& \left.-\la^2c^3\tau xy'x'-2L^2c\dot xy'x'-2c\La\dot x^iy'x^{i\prime}+\la^2c^2\tau x\dot x^ix'x^{i\prime}+2L^2\dot x\dot x^ix'x^{i\prime}+\La\dot x^i\dot x^jx^{i\prime}x^{j\prime}\right\}^\hlf.\non
\eea

We can now apply the decoupling limit discussed in section~\ref{decoupling}. 
In this limit, our parameters scale as follows: 
\bea\label{resc}
g_s(\ap)^\frac{3}{2} &\rr& g_s(\ap)^\frac{3}{2},\non\\
\ap &\rr& \e^{-1}\ap,\\
\la &\rr & \e\la.\non
\eea
In this decoupling limit, our space-time energy $E^-$ is $O(\e)$. We want to consider energies of $O(1)$ in this gauge theory so in~\C{defc},  we choose  $c = 1/\la$ which scales like $\e^{-1}$. Energies with respect to this
choice of world-volume time $\tau$ are finite.  With these choices, we find that
\bea\label{redaction}
S &=& -\frac{1}{g_s\sqrt{\ap}}\int d\tau d\s\left\{\frac{1}{\e^2}+\frac{1}{\e}\dot y-\hlf\left(\dot y^2+\dot x^2+\dot x^i\dot x^i\right)+\sqrt 2\pi\left(x-\tau\dot x\right)F-2\pi^2\La F^2\right.\non\\
&& \left.+\frac{1}{2(\ap)^2}\left(L^2(x')^2+\La\left((y')^2+x^{i\prime}x^{i\prime}\right)\right)+\mathcal{O}(\e)\right\},
\eea
now written in terms of non-scaling quantities.  Note that $\La = \hlf \tau^2 + L^2$ is finite. We can drop the first two terms (a constant and a total $\tau$ derivative), and the omitted higher terms which vanish when $\e\rr 0$ leaving an action which does not scale.

The dimension assignments in~\C{redaction}\ are as follows:  the scalar fields $y$, $x$, $x^i$ have length dimension one, as does $\tau$ while $\s$ is dimensionless.  $A_\tau$ has mass dimension one, while $A_\s$ is dimensionless so that $F$ has uniform mass dimension one. We will rescale these fields to assign canonical dimensions after discussing the non-abelian generalization. Note that the $SO(6)$ symmetry acting on the $x^{i}$ is enhanced to an $SO(7)$ acting on $(x^{i}, y)$.    

\subsection{A non-abelian generalization}

Although we used the DBI action to find the DLCQ description for the $N=1$ case, the natural approach would have been to employ the orbifold description
of the null-brane given by the identification~\C{quotgroup}.  Because the quotient action involves a boost, we will meet some interesting subtleties in
trying to use this approach. 

Let us try to proceed straightforwardly: to describe the theory of $N$ D-branes on the null-brane, we go to the covering space of the 
quotient group action $\Gamma = \Z$, and consider a collection of $(N\times|\G|)\times(N\times|\G|)$ matrices $X^\m$. The  $|\G|=\infty$ label indexes the image branes needed to assure invariance under the quotient action. In fact, these matrices can be viewed as operators on a Hilbert space $\H=\G\otimes\MC^N$.  This picture will be useful below when we want to do a Fourier transformation to a new basis for $\H$.  

There are $\U(N\times|\G|)$ gauge transformations that act on these matrices. We must impose certain constraints both on the matrices $X^\m$ as well as on the gauge transformations to ensure that everything is invariant under the quotient action.

Let us first ignore dynamics and treat the Euclidean D-brane problem, or equivalently the
pure matrix problem. To implement the invariance constraints, let us first define partial matrix elements $X^\m_{mn}=\langle m|X^\m|n\rangle$ which are $N\times N$ Hermitian matrices.  Now the action of the $k^{th}$ quotient group element is easy to understand.  Since the group element acts by the representation $\rho$ where $\rho(k)|n\rangle=|n+k\rangle$, we see that 
$$\left(\rho(k)^\dag X^\m\rho(k) \right)_{mn}=X^\m_{m+k,n+k}.$$  
The constraints on the matrices then become
\bea\label{matconstraints}
X_{m+k,n+k}^{+,i} &=& X_{mn}^{+,i},\non\\
X_{m+k,n+k} &=& X_{mn} + 2\pi k X_{mn}^+,\\
X_{m+k,n+k}^- &=& X_{mn}^- + 2\pi k X_{mn} + 2\pi^2 k^2 X_{mn}^+,\non\\
Z_{m+k,n+k} &=& Z_{mn} + 2\pi kL\dd_{mn}.\non
\eea
The residual gauge transformations are the elements of $\U(N\times|\G|)$ which commute with $\rho(k)$ for all $k$.  Using notation similar to that used above, this simply says that we restrict to unitary matrices $U$ satisfying $U_{m+k,n+k}=U_{mn}$. The action constructed from these matrices is the usual one,
\be\label{action}
S = {1\over 4 g_{s} (\ap)^{2}} \, \Tr \left[ X^\mu, X^\nu\right]^2.
\ee

If we were to study Euclidean D0-branes or D-instantons on the null-brane then we would proceed to solve these pure matrix constraints. 
The solution of these constraints is presented in Appendix~\ref{Dinstantons}. We, however, would like to describe dynamical branes. This involves an identification of world-volume time, $\tau$,
with a space-time coordinate.  Conventional static gauge corresponds to the choice, 
\be
x^0 =  {\tau}. 
\ee 
This choice, however, leads to a theory that involves image branes shifted in time. So in accord with our prior DBI analysis, let us consider the gauge choice
\be
x^+ =  {{\tau}\over \sqrt{2}}. 
\ee
With respect to this choice, each image brane is located at the same point in world-volume time. So we can consider a natural (but by no means unique)
lift of the closed string orbifold identification~\C{quotgroup}\ to dynamical D-branes given by
\bea
X_{m+k,n+k}^{+,i} &=& X_{mn}^{+,i},\label{1constraints}\\
X_{m+k,n+k} &=& X_{mn} + \sqrt{2}\pi k \lambda {\tau } \delta_{mn},\label{2constraints}\\
X_{m+k,n+k}^- &=& X_{mn}^- + 2\pi k \lambda X_{mn} + \sqrt{2}\pi^2 k^2 \lambda^{2}  {\tau} \delta_{mn},\label{3constraints}\\
Z_{m+k,n+k} &=& Z_{mn} + 2\pi kL\dd_{mn}.\non
\eea
We can follow the same procedure described in Appendix~\ref{Dinstantons}\ to transition from 
an infinite-dimensional quantum mechanics system to a $1+1$-dimensional field theory where the fields depend on $(\tau, \T)$ and  $0\le\T\le 1$. Note that $\T$ is dimensionless but $L$ as well as $\tau$ and $X^{\m}$ have length dimension $1$.

Following Appendix~\ref{Dinstantons}\ and performing the Fourier transformation to  $\T$, we obtain
\bea
X^+ &=& x^+,\non\\
X &=& \widetilde{x}+i {\tau \over \sqrt{2}} \lambda \p_\T, \non\\
X^- &=& {\widetilde x}^-+\frac{i}{2} \lambda {\widetilde x}'+i{\widetilde x} \lambda\p_\T- 
{1\over 2 \sqrt{2}}{\tau \lambda^{2}}\p_\T^2, \\
Z &=& z+iL\p_\T,\non\\
X^i &=& x^i,\non
\eea
where small letters represent Hermitian operators that are functions of $(\tau,\T)$, and a prime denotes a derivative with respect to $\T$.  Note that ${\widetilde x}^-$ has been defined so that it is Hermitian.

Gauge transformations, parametrized by a unitary $N\times N$ function $u$ of
the variables $(\T,\tau)$, act in the following way
\bea \label{noninvgt}
&x^+ \rr& ux^+u^\dag,\\
&{\widetilde x} \rr& u{\widetilde x}u^\dag+i{\tau\over \sqrt{2}} \lambda u {u'}^{\dag},\non\\
&{\widetilde x}^- \rr& u{\widetilde x}^-u^\dag+\frac{i}{2} \lambda \left(u{\widetilde x}
{u'}^{\dag}-u'{\widetilde x}u^\dag\right)+{1\over 2\sqrt{2}} \tau \lambda^2 u' {u'}^{\dag},\non\\
&z \rr& uzu^\dag+iLu{u'}^{\dag},\non\\
&x^i \rr& ux^iu^\dag.\non
\eea
Gauge covariant combinations are given by a transformation similar to~\C{hatted}\footnote{This change of variables is singular at the point $L=0$ but is non-singular for any arbitrarily small $L$.}
\bea
& x =& {\widetilde x}-\frac{\tau \lambda}{\sqrt{2} L}z,\\
&{x}^- =& {\widetilde x}^--\frac{\lambda}{2L}\left\{{\widetilde x},z\right\}+\frac{\tau\lambda^2}{2\sqrt{2} L^2}z^2.\non
\eea
In terms of these gauge covariant variables, the above operators are given by
\bea \label{minkop}
X &=& x+ {i\over \sqrt{2}}  {\tau\lambda } D_1,\\
X^- &=& x^-+\frac{i}{2} \lambda \left\{x,D_1\right\}-{1\over 2\sqrt{2}} {\tau\lambda^2} D_1^2,\cr
Z &=& iLD_1\non \eea
where $D_1 = \p_\T - {i\over L} z$. 

There is, however,  a rather crucial difference from the Euclidean case considered in Appendix~\ref{Dinstantons}. We have necessarily treated $X^{+}$ asymmetrically in~\C{1constraints} versus~\C{2constraints} and~\C{3constraints}. On the other hand, $X^{+}$ is treated uniformly in~\C{matconstraints}\ which leads to rather critical cancellations in the resulting Euclidean action~\C{action}. 

In static situations, we transition from D-instantons to D0-branes by making
the replacement $${X^0} \,\rightarrow \, \ap iD^0 = \ap\left( i  \p_\tau + A^0 \right).$$   
However, following this approach using the action~\C{action}\ in this time-dependent case leads to a problematic action precisely because of the asymmetric treatment of $X^+$. Instead we can try to postulate a replacement of $\sqrt 2x^+$ by $\tau+y$ and of $\sqrt{2}x^-$ by $\tau-y$.  This will generate neither gauge-field terms nor kinetic terms for the fields but it does give the remaining interactions. Applying the decoupling limit~\C{resc}\ and sending $\tau \rr \tau/\e$ shows us that $X^-$ given in~\C{minkop}\ collapses to $x^-$. Then setting the rescaled $\lambda=1$ and computing commutators while retaining only terms independent of $\e$ gives
\bea\label{ecomm}
\left[X^+,X\right] &=& \frac{1}{\sqrt 2}\left[y,x\right]-\frac{i\tau}{2}D_1y,\non\\
\left[X^+,X^-\right] &=& 0,\non\\
\left[X^+,Z\right] &=& -\frac{i}{\sqrt 2}LD_1y,\\
\left[X^+,X^i\right] &=& \frac{1}{\sqrt 2}\left[y,x^i\right],\non\\
\left[X,X^-\right] &=& \frac{1}{\sqrt 2}\left[y,x\right]-\frac{i\tau}{2}D_1y,\non\\
\left[X,Z\right] &=& -iLD_1x,\non\\
\left[X,X^i\right] &=& \left[x,x^i\right]+i {\tau \over \sqrt{2}} D_1x^i,\non\\
\left[X^-,Z\right] &=& \frac{iL}{\sqrt 2}D_1y,\non\\
\left[X^-,X^i\right] &=&-\frac{1}{\sqrt 2}\left[y,x^i\right],\non\\
\left[Z,X^i\right] &=& iLD_1x^i,\non\\
\left[X^i,X^j\right] &=& \left[x^i,x^j\right].\non
\eea
where we have defined covariant derivatives, e.g. $D_0x=[D_0,x]$, and a field strength, $F=i[D_0,D_1]$. The action can then be expressed in terms of the commutators~\C{ecomm}\
\bea
S &= {1\over 2 g_{s} (\ap)^{5/2}}\int & d\T d\tau \, \Tr \Big(  -L^2 (D_1x^i)^2 -L^2 (D_1y)^2-L^2 (D_1x)^2 + \hlf \left[ x^i, x^j\right]^2 +\left[ y ,x^i\right]^2 \cr
&& + ( \left[x,x^i\right]+{i\tau\over \sqrt{2}} D_1x^i)^2 +(\left[x,y\right]+{i\tau\over \sqrt{2}}D_1y)^2 +\ldots\Big) .
\eea
The omitted terms involve either time derivatives or gauge-field strengths neither of which are generated by this ansatz. We are also omitting the fermion couplings. Note that, as in the abelian case, $y$ appears symmetrically with the $x^i$ while $x$ is distinguished. 

We can now combine these couplings with our abelian action~\C{redaction}\ to arrive at a conjecture for the complete non-abelian DLCQ description of the null-brane. Since all the parameters are now finite, we can rescale $\T$ by a factor of $\sqrt{\ap}$ to length dimension $1$, rescale all the fields to canonical mass dimension $1$, and define $g^2_{YM} = g_s/\ap$. We can finally set $\ap=1$ for convenience and send $F \rr 2\pi F$. The resulting action is
\bea\label{conjecture}
S &= {1\over 2 g_{YM}^2 }\int & d\T d\tau \, \Tr \Big(  (D_0x^i)^2 + (D_0y)^2+(D_0x)^2-L^2 (D_1x^i)^2 -L^2 (D_1y)^2-L^2 (D_1x)^2 \cr
&& - \sqrt{2} (x - \tau D_0x) F + (L^2+\hlf\tau^2) F^2 +\hlf \left[ x^i, x^j\right]^2 +\left[ y ,x^i\right]^2 \cr
&& + ( \left[x,x^i\right] +{i\tau\over \sqrt{2}} D_1x^i)^2 +(\left[x,y\right]+{i\tau\over \sqrt{2}}D_1y)^2 +{\rm fermions}\Big).
\eea
Note that $(x^i,y)$ are rotated by an $SO(7)$ symmetry and therefore should be combined. 

There is another argument that there are no couplings beyond those seen in~\C{conjecture}.  D0-brane dynamics in flat space should be governed by a matrix quantum mechanics with an action given covariantly by
\be
S=\frac{1}{2g_{YM}^2}\int d\tau\,\Tr\Big(\eta_{\mu\nu}D_0X^\mu D_0X^\nu+\hlf\left[X^\mu,X^\nu\right]\left[X_\mu,X_\nu\right]+{\rm fermions}\Big).
\ee
After substituting~\C{minkop} with $\sqrt 2x^\pm=\tau\pm y$, the kinetic term gives
\bea
(D_0X^\mu)^2 &=& -\frac{1}{\e^2}+(D_0x)^2+(D_0y)^2+(D_0x^i)^2+(L^2+\hlf\tau^2)F^2 \cr
&& +\frac{1}{\sqrt 2}\left(\tau\left\{D_0x,F\right\}-\left\{x,F\right\}\right)+\mathcal{O}(\e).
\eea
After dropping the constant piece and tracing (which now includes an integration over $\T)$, the result agrees precisely with the expression given above in~\C{conjecture}.  The contribution coming from squares of commutators is the same as before and so automatically agrees.

There are a few points worth emphasizing: first, a large gauge transformation along the $\T$ circle simply implements the shift
\be
z \rr z + 2\pi L.
\ee 
In terms of the original variables and their gauge transformation properties given
in~\C{noninvgt}, this large gauge transformation implements the quotient identification. This Lagrangian~\C{conjecture} describes M-theory on the null-brane. However, in agreement with the supergravity solution~\C{stringmetric}, the model is described by a kind of Matrix string theory~\cite{Motl:1997th, Banks:1996my, Dijkgraaf:1997vv}\ near $\tau=0$. On the other hand, as $|\tau|\rr \infty$, fluctuations in $\T$ are suppressed and the model reduces to quantum mechanics. A detailed study of the dynamics of this model will appear elsewhere. 

If we wish to describe perturbative string theory on the null-brane then we need to compactify additional directions in the usual way~\cite{Taylor:1996ik}\ and study higher dimensional generalizations of~\C{conjecture}. This is particularly interesting for type IIB string theory on the null-brane since the conventional IIB Matrix description~\cite{Sethi:1997sw}\ is promoted from a $2+1$ to a $3+1$-dimensional field theory. Lastly, we note that studying D-branes on this kind of quotient space gives a theory that should be closely connected to the dipole models of~\cite{dipoles}, perhaps with a time-dependent dipole. It would be interesting to make this connection precise.

\section*{Acknowledgements}

We would like to thank Ben Craps and Oleg Lunin for helpful discussions. S.~S. would like to thank the organizers of the Third Simons Workshop in Mathematics and Physics at the YITP for providing a stimulating environment during the final stages of this project. 
The work of D.~R. is supported in part by a Sidney Bloomenthal Fellowship.
The work of S.~S. is supported in part by NSF CAREER Grant No. PHY-0094328 and NSF Grant No. PHY-0401814.

\appendix\
\section{Euclidean D0-branes on the Null-brane}
\label{Dinstantons}

In this Appendix, we solve the matrix constraints~\C{matconstraints}\ to obtain an action for Euclidean D0-branes or  D-instantons probing the null-brane. This action has been independently obtained recently in~\cite{Berkooz:2005ym}. We should note that the analytic continuation to Euclidean space needed to 
describe D-instanton configurations is not straightforward for the null-brane. It is unclear whether
physical amplitudes in type II string theory can really receive quantum corrections from these kinds of D-instantons. For us, however, the solution of the pure matrix problem is an intermediate step on the road to describing dynamical D-branes.  

We wish to solve the pure matrix constraints~\C{matconstraints}. Solving these constraints allows one to express the matrices $X^\m_{mn}$ in terms of just $X^\m_{m0}$.  These latter matrices are the residual degrees of freedom.  A more convenient picture is obtained by changing basis, from $|n\rangle$ to
\be
|\T\rangle=\sum_ke^{2\pi ik\T}|k\rangle,
\ee
where now $0\le\T\le 1$.  The inner product is $\langle\T'|\T\rangle=\dd(\T-\T')$, and the identity can be written as 
$${\operatorname{Id}}=\int d\T |\T\rangle\langle\T|.$$  
By rewriting the probe theory in this way, our matrices become functions of a single periodic variable $\T$.  In other words, we have effectively implemented a T-duality along the quotient direction to obtain a theory of D-instantons in the T-dual geometry.  This is very much along the lines used in~\cite{Taylor:1996ik}\ to describe circle compactifications. 

Let us define matrix elements with respect to this new basis, 
\be
X^\m(\T,\T')\equiv\langle\T|X^\m|\T'\rangle = \sum_{m,n} e^{2\pi i(n\T'-m\T)}
X_{mn}^\m. 
\ee
The solution to the $X^\m$ constraints is then given by, 
\bea
X^{+,i}(\T,\T') &=& x^{+,i}(\T) \dd(\T-\T'),\non\\
X(\T,\T') &=& \left[x(\T)+ix^+(\T)\p_\T \right]\dd(\T-\T'),\\
X^-(\T,\T') &=& \left[x^-(\T)+ix(\T)\p_\T-\hlf x^+(\T)\p_\T^2\right]\dd(\T-\T'),\non\\
Z(\T,\T') &=& \left[z(\T)+iL\p_\T\right]\dd(\T-\T'),\non
\eea
where for each $X^\m$ we have defined
\be
x^\m(\T)\equiv\sum_{k}e^{-2\pi ik\T}X_{k0}^\m.
\ee
Each of these operators is local in $\T$ in the sense that they can be written
as $A(\T,\T')=\hat A(\T)\dd(\T-\T')$ for some operator $\hat A(\T)$.  For any
two operators $A$, $B$ which are local in this sense it is easy to check that
$$(A\cdot B)(\T,\T')=\hat A(\T)\cdot\hat B(\T)\cdot\dd(\T-\T'),$$ 
so we can multiply operators locally.  We will also drop any hats, since it will be
clear from the number of parameters which object we mean.

There is a problem in this setup, however; the $N\times N$ matrices $x^\m(\T)$ are not necessarily Hermitian.  Indeed, as an example consider
\bea
\left[x(\T)\right]^\dag &=& \left[\sum_ke^{-2\pi ik\T}X_{k0}\right]^\dag=\sum_ke^{2\pi ik\T}X_{0k}^\dag=\sum_ke^{2\pi ik\T}\left[X_{-k,0}^\dag+2\pi k{X^+_{-k,0}}^\dag\right]\non\\
&=& x(\T)-i x^+(\T)',
\eea
where a prime represents differentiation with respect to $\T$.  To fix this problem we can define operators
\bea
\tilde x(\T) &=& x(\T)-\frac{i}{2}x^+(\T)',\\
\tilde x^-(\T) &=& x^-(\T)-\frac{i}{2} x(\T)'-\frac{1}{4} x^+(\T)'',\non
\eea
which are Hermitian.

The gauge transformations acting on our fields are
\bea
&x^{+} \rr ux^{+}u^\dag,&\\
&x^{i} \rr ux^{i}u^\dag,&\non\\
&\tilde x \rr u\tilde xu^\dag+\frac{i}{2}\left(ux^+{u'}^\dag - u' x^+ u^\dag\right),&\non\\
&\tilde x^- \rr u\tilde x^-u^\dag+\frac{i}{2}\left(u\tilde x  {u'}^\dag- u'\tilde xu^\dag\right)-\frac{1}{4}\left(u x^+ {u''}^\dag + u {x^+}' {u'}^\dag+ u' {x^+}' u^\dag+ {u''} x^+u^\dag\right),&\non\\
&z \rr uzu^\dag+iLu {u'}^\dag &\non
\eea
where $u=u(\theta)$ is a unitary $N\times N$ matrix.
In terms of these variables we can then compute the commutators.  We find (dropping tildes and delta functions)
\bea
&\left[X^+,X\right] = & [x^+,x]-\frac{i}{2}\{x^+,{x^+}'\}, \\
&\left[X^+,X^-\right] =& [x^+,x^-]+\frac{i}{2}\left([x^+, x']-2x {x^+}' \right)+\hlf\left(({x^+}')^2+x^+ {x^+}'' \right)\non\\
 &&+i[x^+,x]\p+\hlf\{x^+, {x^+}'\}\p,\non\\
&\left[X^+,Z\right] =& [x^+,z]-iL {x^+}',\non\\
&\left[X^+,X^i\right] =& [x^+,x^i],\non\\
&\left[X,X^-\right] =& [x,x^-]-\frac{i}{2}\{x, x' \}+\frac{1}{4}\left(\{ {x^+}', x'\}+2x {x^+}'' \right)
\non\\
 &&+\frac{i}{2}\left([ {x^+}' ,x^-]+2x^+{x^-}' \right)+\frac{i}{4}\left( {x^+}' {x^+}'' + x^+{x^+}'''\right)\non\\
 &&-\hlf\left([x^+,x']-2x {x^+}'\right)\p+i[x^+,x^-]\p+\frac{i}{2}\left(( {x^+}')^2+x^+
 {x^+}''\right)\p \non\\
 &&-\hlf[x^+,x]\p^2+\frac{i}{4}\{x^+,{x^+}' \}\p^2,\non \eea
 \bea
&\left[X,Z\right] =& [x,z]+\frac{i}{2}\left([ {x^+}',z]+2x^+z'\right)-iL x'+\hlf L {x^+}''
+i[x^+,z]\p+L {x^+}' \p, \non\\
&\left[X,X^i\right] =& [x,x^i]+\frac{i}{2}\left([ {x^+}',x^i]+2x^+{x^i}'\right)+i[x^+,x^i]\p, \non\\
&\left[X^-,Z\right] =& [x^-,z]-iL {x^-}' +\hlf L x''+\frac{i}{2}\left([x',z]+2x z'\right)-\hlf\left( {x^+}' z'+x^+ z'' \right)+L x' \p \non\\
&& +\frac{i}{2}L {x^+}''\p+i[x,z]\p-\hlf\left([ {x^+}',z]+2x^+z'\right)\p+\frac{i}{2}L 
{x^+}' \p^2-\hlf[x^+,z]\p^2, \non\\
&\left[X^-,X^i\right] =& [x^-,x^i]+\frac{i}{2}\left([x',x^i]+2x {x^i}' \right)-\hlf\left(
{x^+}' {x^i}'+x^+{x^i}''\right) \non\\
&& +i[x,x^i]\p-\hlf\left([ {x^+}',x^i]+2x^+{x^i}' \right)\p-\hlf[x^+,x^i]\p^2, \non\\
&\left[Z,X^i\right] =& [z,x^i]+iL {x^i}', \non\\
&\left[X^i,X^j\right] =& [x^i,x^j], \non
\eea
where all of the $x^\m$ are functions of $\T$, primes represent differentiation with respect to $\T$, and $\p={\p \over \p\T}$. 
The action for the pure matrix theory is then given by a trace of commutators squared,
\be
S=\frac{1}{4g_{s}(\ap)^{2}} \int d\T \, \Tr\left[X^\m,X^\n\right]^2,
\ee
and notably involves higher derivative interactions. 

This action is quite complicated in the non-abelian case. However,
the result simplifies immensely for the abelian case since all commutators drop out.  The result is
\bea
S &=& \frac{1}{2g_{s}(\ap)^{2}}\int d\T\left\{\left(L^2+(x^+)^2\right)\left(2x^{+\prime}x^{-\prime}-\left(x^{i\prime}\right)^2\right)-L^2\left(x'\right)^2-2x^+xx^{+\prime}x'+x^2\left(x^{+\prime}\right)^2\vphantom{\frac{1}{4}}\right.\non\\
&& \left.+2L\left(x^+x'-xx^{+\prime}\right)z'-(x^+)^2\left(z'\right)^2-\frac{1}{4}L^2\left(x^{+\prime\prime}\right)^2+x^+\left(x^{+\prime}\right)^2x^{+\prime\prime}+\hlf\left(x^+\right)^2x^{+\prime}x^{+\prime\prime\prime}\right.\non\\
&& \left.+\frac{1}{4}\left(x^{+\prime}\right)^4+\frac{1}{4}\left(x^+\right)^2\left(x^{+\prime\prime}\right)^2\right\}.
\eea
This action is gauge invariant, as can be seen by switching to gauge invariant coordinates
\bea
\hat x &=& x-L^{-1}zx^+,\\
\hat x^- &=& x^--L^{-1}zx+\hlf L^{-2}z^2x^+.\non
\eea
In terms of these variables the action is
\bea
S &=& \frac{1}{2g_{s}(\ap)^{2}}\int d\T\left\{\left(L^2+(x^+)^2\right)\left(2x^{+\prime}\hat x^{-\prime}-\left(x^{i\prime}\right)^2\right)-L^2\left(\hat x'\right)^2-2x^+\hat xx^{+\prime}\hat x'+\hat x^2\left(x^{+\prime}\right)^2\vphantom{\frac{1}{4}}\right.\non\\
&& \left.-\frac{1}{4}L^2\left(x^{+\prime\prime}\right)^2+x^+\left(x^{+\prime}\right)^2x^{+\prime\prime}+\hlf\left(x^+\right)^2x^{+\prime}x^{+\prime\prime\prime}+\frac{1}{4}\left(x^{+\prime}\right)^4+\frac{1}{4}\left(x^+\right)^2\left(x^{+\prime\prime}\right)^2\right\},
\eea
which is manifestly gauge invariant since the only charged field, $z$, drops out.  In fact the two derivative terms in this action are exactly what one would obtain using DBI for the case of a Euclidean D0-brane wrapping the T-dual geometry.

%\newpage
%\bibliographystyle{amsunsrt-es}
%\bibliography{myrefs}
%\begin{thebibliography}{10}
%\bibliographystyle{utphys}
%\bibliography{myrefs}

\providecommand{\href}[2]{#2}\begingroup\raggedright\endgroup
\end{document}